\documentclass{iopart}
\usepackage{amssymb,mathptmx,times,graphicx,rangcite}
\catcode`\@ 11
\renewcommand\paragraph{\@startsection{paragraph}{4}{\z@}%
  {1ex \@plus1ex \@minus.2ex}{-1em}{\reset@font\normalsize\itshape}}
\catcode`\@ 12
\advance\parskip 4pt
\advance\headsep 2pt
\advance\leftmargini 0.2em
\eqnobysec
\DeclareMathAlphabet\mathpzc{OT1}{pzc}{m}{it}
\let\mathcal=\mathpzc
\def\[{\begin{equation}}
\def\]{\end{equation}}
\def\bse{\numparts}
\def\ese{\endnumparts}
\def\bea{\begin{eqnarray}}
\def\eea{\end{eqnarray}}

\def\d{{\rm d}}
\def\e{{\rm e}}
\def\F{{\cal F}}

\def\half{{\textstyle\frac12}}
\let\trueint=\int
\def\int{\mathop{\textstyle\trueint}\limits}
\def\iint{\mathop{\textstyle\int\kern-0.4em\int}\limits}
\def\iiint{\mathop{\textstyle\int\kern-0.4em\int\kern-0.4em\int}\limits}
\newcommand\partialderiv[3][]{\frac{\partial^{#1}#2}{\partial {#3}^{#1}}}
\newcommand\deriv[3][]{\frac{d^{#1}#2}{d{#3}^{#1}}}
\def\Laplacian{\nabla^2_{\!\!\@x\,}}
\def\Re{\mathop{\rm Re}\nolimits}
\def\Im{\mathop{\rm Im}\nolimits}
\def\re{{\rm re}}
\def\im{{\rm im}}
\def\Real{{\mathbb{R}}}
\def\Complex{{\mathbb{C}}}

\def\sech{\mathop{\rm sech}\nolimits}
\def\sinc{\mathop{\rm sinc}\nolimits}

\def\ci{\mathop{\rm ci}\nolimits}
\let\^=\hat
\let\==\bar
\let\@=\mathbf
\def\_#1{_{\mathrm{#1}}}
\let\~=\mathsf
\def\O#1{^{(#1)}}
\def\nosp{\kern-0.60em}
\let\<=\langle
\let\>=\rangle
\def\numparts{\refstepcounter{equation}%
     \setcounter{eqnval}{\value{equation}}%
     \setcounter{equation}{0}%
     \def\theequation{\arabic{section}.\arabic{eqnval}{\it\alph{equation}}}}
\def\endnumparts{\def\theequation{\arabic{section}.\arabic{equation}}%
     \setcounter{equation}{\value{eqnval}}}

\begin{document}
\thispagestyle{empty}
\title[The DMGLE and its application to femtosecond lasers]{The dispersion-managed Ginzburg-Landau equation and its application to femtosecond lasers}
\author{Gino Biondini}
\address{State University of New York at Buffalo, Department of Mathematics, Buffalo, NY 14260}

\begin{abstract}
The complex Ginzburg-Landau equation is a universal model which
has been used extensively to describe various non-equilibrium phenomena. 
In the context of lasers, it models the dynamics of a pulse
by averaging over the effects that take place inside the cavity.
Ti:sapphire femtosecond lasers, however, produce pulses that
undergo significant changes in different parts of the cavity 
during each round-trip. 
The dynamics of such pulses is therefore not adequately described 
by an average model that does not take such changes into account.
The purpose of this work is severalfold.  First we introduce the 
dispersion-managed Ginzburg-Landau equation (DMGLE) as an average 
model that describes the long-term dynamics of systems characterized
by rapid variations of dispersion, nonlinearity and gain in a general
setting, and we study the properties of the equation.
We then explain how in particular the DMGLE arises for Ti:sapphire
femtosecond lasers and we characterize its solutions.
In particular, we show that, for moderate values of the gain/loss
parameters, the solutions of the DMGLE are well approximated by
those of the dispersion-managed nonlinear Schr\"odinger equation (DMNLSE),
and the main effect of gain and loss dynamics is simply to select one 
among the one-parameter family of solutions of the DMNLSE.
\par\medskip\noindent\today
\par\kern-2\bigskipamount
\end{abstract}


\section{Introduction}
\label{s:intro}

The complex Ginzburg-Landau equation (CGLE) is a universal model 
that governs the non-equilibrium dynamics of weakly nonlinear systems 
in the presence of gain, saturation, as well as linear and
nonlinear dispersion,
and describes a large variety of physical phenomena, from
nonlinear waves to second-order phase transitions, superconductivity
superfluidity, Bose-Einstein condensation, liquid crystals and
strings in field theory~\cite{RMP74p99,RMP65p851}.
In particular, equations of CGLE type have often been used as a model 
for the description of mode-locked lasers 
\cite{
JSTQE6p1173,JQE28p2086,SIAMREV48p629,PRL73p2978,PHYSD83p478,PRA44v4712,PRA48p1573}.
The most well-known example in this context is referred to as the 
master equation of passive mode-locking~\cite{JSTQE6p1173,SIAMREV48p629}.

Lasers have a number of important scientific and 
technological applications, of course.
In particular, 
the latest generation of Ti:sapphire femtosecond lasers 
\cite{RMP75p325,YeCundiff}
could be used in spectroscopy, frequency metrology, and 
optical atomic clocks \cite{Science293p825,Science288p635,Nature396p239,Nature416p233}. 
One of the features of these lasers is the 
presence of large variations of dispersion and nonlinearity 
inside the cavity~\cite{JOSAB16p1999}.
The two phenomena are called respectively dispersion management and 
nonlinearity management by analogy with optical fiber communications.
Because of these variations,
another mathematical tool which has been recently used in studies of
femtosecond lasers
is the dispersion-managed nonlinear Schr\"odinger equation (DMNLSE), 
which is a universal model that describes the long-term dynamics 
of weakly nonlinear dispersive pulses subject to large, 
periodic variations of the nonlinear and/or dispersion 
coefficients~\cite{OL23p1668,OL21p327}.\break
It should be clear, however, 
that the CGLE and DMNLSE are both inadequate descriptions of
Ti:sapphire lasers.
On one hand, the CGLE is derived under the assumption that 
the pulse changes per round-trip inside the cavity are small, 
a condition which is grossly violated in femtosecond lasers.  
As a result, the CGLE cannot provide an accurate description 
of the dynamics of such lasers.
On the other hand, 
gain/loss effects are obviously significant in \textit{any} laser, 
so the DMNLSE cannnot be a fully accurate description of a laser.

In this work we address both of the above issues.
Starting from a multi-dimensional perturbed CGLE with 
time-dependent coefficients,
we derive a new equation --- which we call the
dispersion-managed Ginzburg-Landau equation (DMGLE) --- 
that describes the long-term dynamics of weakly nonlinear systems
subject to rapidly varying dispersion, nonlinearity and gain/loss.
We then study the properties of this equation and we use it to investigate 
the behavior of dispersion-managed solitons in femtosecond lasers.
The outline of this work is the following:
In section~\ref{s:dmgle} we obtain the DMGLE in a general setting
as the equation which describes the average dynamics of systems
governed by the CGLE with large  and rapidly varying 
coefficients.
In section~\ref{s:lasers} we then show how in particular the DMGLE 
arises in the description of Ti:sapphire femtosecond lasers.
In section~\ref{s:symmetries} we study the symmetries,
rate equations and linear modes of the DMGLE,
and in section~\ref{s:solutions} we discuss solitary-wave 
solutions and their properties.
Section~\ref{s:conclusions} concludes with some final remarks.

\section{From the CGLE to the DMGLE}
\label{s:dmgle}

Here we obtain the DMGLE in a general setting as the equation which 
describes the long-time dynamics of multidimensional systems governed 
by the CGLE with large, 
time-dependent and rapidly varying coefficients.
We start from a perturbed cubic CGLE in dimensionless form
with time-dependent coefficients
\[
i\partialderiv qt+\frac12 p(t/t_a,t) \Laplacian q
  + n(t/t_a,t)|q|^2q  = i g(t/t_a,t)q\,,
\label{e:CGLE+DM}
\]
where $q=q(\@x,t)\in\Complex$,\,
$\@x=(x_1,\dots,x_N)\in\Real^N\!$,\, $t\in\Real$, 
$\Laplacian$ is the Laplacian operator, 
and the coefficients $p$ and $n$ are complex:
$p(\cdot)= p_\re(\cdot)+ip_\im(\cdot)$ etc.
In the context of mode-locked lasers, these quantities have the following
physical interpretation:
\par\kern-\medskipamount
\begin{itemize}
\itemsep0pt\parsep0pt
\item
$p_\re(t/t_a,t)$ quantifies chromatic dispersion,
\item
$p_\im(t/t_a,t)$ quantifies spectral filtering or band-limited gain,
\item
$n_\re(t/t_a,t)$ quantifies nonlinear interactions mediated by the Kerr effect,
\item
$n_\im(t/t_a,t)$ quantifies gain saturation,
\item
$g_\re(t/t_a,t)$ quantifies linear gain
\end{itemize}
\par\kern-\medskipamount
\noindent
(see section~\ref{s:lasers} for a more detailed discussion).
Here $t_a$ is a short temporal scale characteristic of the problem, 
which will be defined more precisely below.
Without loss of generality, the function $g(t/t_a)$ can be taken to be real,
since an imaginary component of $g(t/t_a)$ can always be eliminated
by defining a rescaled field via the phase 
$\exp\big[i\int g_\im(t/t_a)\,\d t\big]$.
But the same statement does not apply to $p(t/t_a)$ and $n(t/t_a)$,
of course.

Solutions of~\eref{e:CGLE+DM}
with complex coefficients have been extensively studied;
e.g., see Refs.~\cite{AkhmedievAnkiewicz,RMP74p99} and references therein.
These studies treat the standard CGLE, namely, the case in which 
all the coefficients are constant.
Solutions of~\eref{e:CGLE+DM} with large, 
time-dependent and rapidly varing coefficients, on the other hand,
can be very different from those of the standard CGLE,
and characterizing their behavior is a nontrivial task.
Moreover, to the best of our knowledge, 
such problem has not been addressed in the literature, despite its
theoretical and practical relevance.
Our goal is precisely to elucidate the dynamics of such solutions.
We do so by employing a multiple-scale analysis similar to
that used to study optical fiber communication systems 
in the presence of dispersion management~\cite{OL23p1668,OL22p985}.

\paragraph{Multiple scales expansion and the DMGLE.}

If the temporal variations of the coefficients in~\eref{e:CGLE+DM}
are not too large, one can employ a standard
multiple-scale perturbation expansion,
as was done for the NLSE in~\cite{OL22p985}.
The result is that, to leading-order,
the pulse satisfies a CGLE with constant coefficients that are simply 
the averages of those appearing in~\eref{e:CGLE+DM},
Higher-order terms then give corrections to this leading-order behavior.
(Of course different starting models can give rise to different scenarios,
e.g., see Refs.~\cite{Malomed2005,PRL91p240201}.)
When the coefficients in~\eref{e:CGLE+DM} exhibit large temporal variations, 
however, the situation is not as simple, 
since the standard perturbation expansion breaks down, 
as it does for the NLSE~\cite{OL22p985}.
It is this latter case that is of interest here.
There are three key assumptions for our analysis:
\begin{enumerate}
\item
All of the coefficients in~\eref{e:CGLE+DM} are periodic 
with respect to their first argument.
That is,
$p(\zeta+1,t)= p(\zeta,t)$ $\forall\zeta\in\Real$ and similarly for 
$n(\,\cdot\,)$ and $g(\,\cdot\,)$.
\item
The coefficients in~\eref{e:CGLE+DM} are rapidly varying.
That is, the dimensionless period is $t_a\ll1$.
\item
The function $p(\,\cdot\,)$ can be decomposed into the sum of two terms,
describing respectively large periodic zero-mean oscillations and 
an $O(1)$ residual.  Namely, 
\bse
\bea
p(t/t_a,t)= \frac1{t_a}\,\Delta p(t/t_a) + p\_{res}(t/t_a,t)\,,
\label{e:mscaling}
\\
\noalign{\hglue\leftmargini with}
\int_0^{t_a}\Delta p(t/t_a)\,\d t=0\,.
\eea
\ese
\end{enumerate}
Note that both $\Delta p$ and $p\_{res}$ can have both a real and an
imaginary component.
(Indeed, this will be the case in section~\ref{s:lasers}.)
Note also that 
the double time dependence of $p(\cdot)$, $n(\cdot)$ and $g(\cdot)$ 
in~\eref{e:CGLE+DM}
allows one to capture cases in which the corresponding physical parameters
have both a fast and a slowly varying component with respect to time.
Even though the simplest case is when all of the
coefficients average to a constant 
(as in optical fiber communications~\cite{OL23p1668}),
the more general situation does not introduce significant complications 
in the method.

With the above assumptions, we introduce the fast temporal scale 
$\zeta= t/t_a$ and look for solutions in the form
\[
q(\@x,\zeta,t)= q\O0(\@x,\zeta,t)+ t_a\,q\O1(\@x,\zeta,t) + O(t_a^2)\,.
\label{e:msansatz}
\]
It is convenient to denote by $\=f(\@x,t)$ the average of 
a function $f(\@x,t/t_a,t)$ over one map period, namely:
\[
\=f(\@x,t)= \int_0^1 f(\@x,\zeta,t)\,\d\zeta= 
  \frac1{t_a}\int_0^{t_a} f(\@x,t'/t_a,t)\,\d t'\,.
\]
We then substitute the ansatz~\eref{e:msansatz} into 
the CGLE~\eref{e:CGLE+DM} 
and solve the resulting perturbation expansion following 
standard procedures.
%
At leading order in the expansion we have the following linear problem:
\[
i\partialderiv{q\O0\nosp}\zeta
  + \frac12 \Delta p(\zeta)\,\Laplacian q\O0= 0\,.
\label{e:mslinearproblem}
\]
The solution of~\eref{e:mslinearproblem} is of course trivially
obtained using Fourier transforms.
We denote Fourier transforms by a circumflex accent throughout, 
with the transform pair defined as
\bse
\bea
\^f(\@k)=
\F[f(\@x)]= \int \e^{i\@k\cdot\@x}f(\@x)\,(\d\@x)\,,
\label{e:FT}
\\
f(\@x)=
\F^{-1}[\^f(\@k)]= \frac1{(2\pi)^N} 
  \int \e^{-i\@k\cdot\@x}\^f(\@k)\,(\d\@k)\,,
\label{e:IFT}
\eea
\ese
where $(\d\@x)=\d x_1\cdots\d x_N$ is the volume element in $\Real^N$.
(Integrals are complete --- meaning over all of $\Real^N$ --- 
throughout this work unless otherwise noted.)
With these conventions, 
the general solution of~\eref{e:mslinearproblem} is 
\bea
\^q\O0(\@k,\zeta,t)= \^\chi(\@k,\Xi(\zeta))\,\^u(\@k,t)\,,\qquad
\^\chi(\@k,\xi)= \e^{-i\xi k^2/2}\,,
\label{e:solndecomposition} 
\\
\noalign{\noindent where $k^2=\@k\cdot\@k$,} 
\Xi(\zeta)= 
  \Xi_o + \int_0^\zeta \Delta p(\zeta')\,\d\zeta'
\label{e:mdef}
\eea
and $\Xi_o$ is an arbitrary integration constant.
Equation~\eref{e:solndecomposition} separates the fast dynamics 
of the pulse from the slow dynamics:
The exponential factor in~\eref{e:solndecomposition} 
accounts for the rapid ``breathing'' (periodic compression/expansion)
of the pulse, while the
slowly varying envelope $\^u(\@k,t)$ encodes the information
about the core pulse shape.
Inverting the Fourier transforms we get the leading-order solution 
in the spatial domain,
\[
q\O0(\@x,\zeta,t)= \chi(\@x,\Xi(\zeta)) * u(\@x,t)\,,\qquad
\chi(\@x,\xi)= \frac{\e^{ix^2/2\xi}\nosp}{\sqrt{(2\pi i)^N\xi}}\,,
\]
where $x^2=\@x\cdot\@x$ and $\sqrt{z}$ is taken on the principal branch 
(with $|\arg(z)|<\pi$),
and where the asterisk denotes the convolution integral,
defined as
\begin{equation}
(f*g)(\@x)=
\int f(\@x')g(\@x-\@x')\,(\d\@x')\,,
\end{equation}
Above and below we used the convolution theorem,
stating that 
$\F^{-1}[\^f\^g]= f*g$\ and
$\F[fg]= \^f*\^g\,/(2\pi)^N$\,.
Note that $\^\chi(\@k,0)=1$ and $\chi(\@x,0)=\delta(\@x)$,
where $\delta(\@x)$ is the $N$-dimensional Dirac delta.
Therefore, at those points $\zeta$ for which $\Xi(\zeta)=0$,
the solution $q\O0(\@x,\zeta,t)$ coincides with $u(\@x,t)$.
Note also that, since $\chi(\@x,\xi)$ is even in $\@x$, 
the parity of $q\O0(\@x,\cdot,\cdot)$ 
is determined by that of $u(\@x,\cdot)$. 
That is, if $u(\@x,t)$ is even with respect to $\@x$, 
$q\O0(\@x,\zeta,t)$ is also even; 
vice versa, if $u(\@x,t)$ is odd $q\O0(\@x,\zeta,t)$ is also odd.

The function $u(\@x,t)$ is arbitrary at this stage, and as usual 
must be determined at higher order in the expansion.
At $O(1)$ we have the following forced linear problem:
\bea
i\partialderiv{q\O1\nosp}\zeta
  + \frac12\Delta p(\zeta)\,\Laplacian q\O1=
\nonumber
\\
\kern4em{ } 
  - \bigg[ i\partialderiv{q\O0\nosp}t + \frac12p\_{res}(\zeta,t)\,\Laplacian q\O0
      + n(\zeta,t)|q\O0|^2q\O0 - ig(\zeta,t)q\O0 \bigg].
\nonumber
\\{ }
\label{e:orderone}
\eea
The solution of~\eref{e:orderone} can also be easily obtained 
using Fourier transforms.
To avoid secularities, however, a Fredholm solvability condition 
must be satisfied (which as usual expresses the orthogonality of the forcing 
to the solutions of the homogeneous problem).
After tedious but straightforward algebra, this condition yields
\bea
i \partialderiv ut + \frac12 \=p(t)\,\Laplacian u + \iint
  u_{(\@x+\@x')}u_{(\@x+\@x'')}u^*_{(\@x+\@x'+\@x'')}K_{(\@x'\cdot\@x'')}\,
    (\d\@x')(\d\@x'')  = i\= g(t)\,u\,,
\nonumber
\\*[-0.5ex]
\label{e:DMGLEo}
\eea
or, equivalently, in the Fourier domain,
\bea
i\partialderiv{\^u}t - \frac12 \=p(t)\,k^2\^u + \iint
  \^u_{(\@k+\@k')}\^u_{(\@k+\@k'')}\^u^*_{(\@k+\@k'+\@k'')}
    \^K_{(\@k'\cdot\@k'')}\,(\d\@k')(\d\@k'')  =
     i\= g(t)\,\^u,
\nonumber
\\*[-0.5ex]
\label{e:DMGLEf}
\eea
where for brevity we have introduced the shorthand notation 
$u_{(\@x)}=u(\@x,t)$ and $\^u_{(\@k)}= \^u(\@k,t)$.
(That is, subscripts in parenthesis denote functional dependence,
\textit{not} partial differentiation.)
The superscript * denotes complex conjugation throughout,
and the integration kernels (which are in general complex and time-dependent) 
are given by
\bse
\bea
\^K(\xi,t)= \frac1{(2\pi)^{2N}}
  \int_0^1 n(\zeta,t)\,\e^{i\Xi(\zeta)\xi}\,\d\zeta\,
\label{e:Khatdef}
\\
\noalign{\noindent and}
K(\@x'\cdot\@x'',t)= (2\pi)^{2N}\F^{-1}[\^K(\@k'\cdot\@k'',t)]\,.
\label{e:Kdef}
\eea
\ese
[Again, a straightforward calculation shows that,
as in the one-dimensional DMNLSE~\cite{OL23p1668}, 
$K(\cdot,t)$ depends only on the dot product 
$\@x'\cdot\@x''$, not on $\@x'$ and $\@x''$ separately.
Note also that $\=p(t)=\=p\_{res}(t)$.]
We refer to~\eref{e:DMGLEo} [or equivalently~\eref{e:DMGLEf}] as the 
dispersion-managed Ginzburg-Landau equation, or DMGLE.

\paragraph{Remarks.}

Several comments are now in order:
\begin{itemize}
\item
If $\Delta p(\zeta)=0$, then choosing $\Xi_o=0$ one has
$\Xi(\zeta)= 0$, 
which implies $\^K(\xi,t)= 1/(2\pi)^{2N}$
and $K(\@x'\cdot\@x'',t)= \delta(\@x'-\@x'')$\,.
Then~\eref{e:DMGLEo} and~\eref{e:DMGLEf} reduce to the usual CGLE 
with constant coefficients written in the spatial and Fourier domains.
\item
If $p_\im(\zeta,t)=n_\im(\zeta,t)=0$, equation~\eref{e:DMGLEo} reduces
to the multidimensional extension of DMNLSE~\cite{OL23p1668,OL21p327}, 
perturbed by the addition of linear gain and loss when $\=g(t)\ne0$.
Then, if $\Delta p(\zeta)=0$, equation~\eref{e:DMGLEo} 
further reduces to the NLSE with constant coefficients,
again possibly with gain and loss. 
\item 
Since the NLSE, the CGLE and the DMNLSE arise in many different 
physical situations, one can expect that the DMGLE will also have 
a wide range of applicability.
Indeed, the above formulation is fairly general, and it allows one
to study a variety of different scenarios,
including spectral filtering and nonlinear loss and gain.
\item
Equations of NLS type or CGLE type with nonlocal nonlinearities arise
in various physical situations.
In particular, apart from optical fiber communications, 
they also appear in water waves, in which case they are 
known to be integrable~\cite{PLA197p401}.
\item
Like the DMNLSE (its counterpart for conservative systems),
the DMGLE is a reduced model that retains 
the essential features of dispersion-managed systems while bypassing the 
complicated dynamics that take place within each period.
As such, it should prove to be a useful model to investigate 
the long-time behavior of nonconservative dispersion-managed systems,
like the DMNLSE did for conservative ones. 
\item 
Since $q(x,\zeta_*,t)= u(x,t)$ whenever $\Xi(\zeta_*)=0$,
the value of $\Xi_o$ simply dictates at which points inside the map
the full solution of~\eref{e:CGLE+DM} coincides with 
its core pulse shape.
At the same time, however, the choice of $\Xi_o$ appears in 
$\^K(\xi,t)$ and its transform, 
and therefore can sometimes be exploited to make them simpler.
(For example, the corresponding kernels for the DMNLSE in the lossless case
can be made real~\cite{OL23p1668}.)
\item 
If $n(\zeta,t)\ne{}$const, the kernels $\^K(\xi,t)$ and $K(\xi,t)$ 
can have a nonzero imaginary part even when $\Delta p_\im(\zeta,t)=0$.
Indeed, this is the case in optical fiber communication systems 
with loss compensated by periodically spaced amplifiers,
when one takes into account that the effective nonlinear coefficient
in the NLSE is distance-dependent due to the periodic power variations
originated by the loss/amplification cycle. 
\item 
A multiple-scale averaging similar to the one described above 
can also be performed for the quintic Ginzburg-Landau equation.
The result of such an averaging is that the quintic term in the 
original equation produces a quintic term in the DMGLE 
with a four-dimensional convolution integral and a corresponding
integration kernel.
Such a situation will not be considered here, however.
\end{itemize}
Finally, we reiterate that $\Delta p(\zeta)$ need not be real
for the purposes of the multiple-scale expansion:
the derivation applies without change when $\Delta p_\im(\zeta)\ne0$.
Of course, some of the conclusions are different:
$\Xi(\zeta)$ acquires an imaginary part, and
this has two main consequences:
(i)~it leads to a periodic amplification/suppression
of high-wavenumber components in the fast dynamics;
(ii)~it leads to different kernels in the DMGLE.
While the first of these consequences can be considered to be minor
(since it has no effect on the long-term behavior of solutions),
the second can in principle produce different and interesting
phenomena.
%
%

\section{The DMGLE in femtosecond lasers}
\label{s:lasers}

\begin{figure}[t!]
\medskip
\centerline{\includegraphics[width=0.6\textwidth]{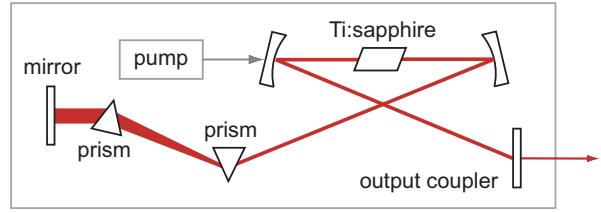}}
\caption{Schematic diagram of a prototypical Ti:sapphire femtosecond laser.
See text for a detailed description.}
\label{f:laser}
\end{figure}

We now discuss how the DMGLE appears in the description of Ti:sapphire 
lasers.
\Fref{f:laser} shows a diagram of a typical
experimental setup.
The system consists of a continuous-wave (CW) pump, 
a Titanium-doped sapphire crystal and 
a set of prisms and/or mirrors.
The crystal, which constitutes the nonlinear medium, is characterized by 
a Kerr-type nonlinear response as well as by gain and gain saturation, 
and has a large normal group velocity dispersion.
[That is, for CW waves whose electric field is written as 
$\Re(E\,\e^{ikz-\omega t})$,
where $k(\omega)$ represents the linear dispersion relation it is 
$k''=\partial^2k/\partial \omega^2<0$.]
The prisms and mirrors, on the other hand, are especially designed 
to have a large anomalous dispersion to compensate that of the 
crystal.
The output of such a system is a periodic stream of optical pulses 
with a typical duration of about 10\,fs,
a spectrum typically centered at 830\,nm and 70\,nm wide,
and with a typical repetition rate of 90\,MHz
(e.g., see Ref.~\cite{Science288p635}).

\paragraph{Laser model.}

The evolution of a quasi-monochromatic optical pulse in a Kerr-type medium
is usually well described by the NLSE, possibly with varying coefficients
\cite{Agrawal,JOSAB16p1999,PRL94p243904}.  
%
To accurately capture the behavior of pulses in a laser, however,
it is obviously necessary take into account the gain and
loss dynamics \cite{JOSAB16p1999,
OE15p6677}.
In Ti:sapphire lasers,
the gain dynamics (including linear gain and gain saturation) takes place 
inside the crystal, while the losses are concentrated in the reflecting 
mirror(s) and the output coupler.
Since the precise dependence of the gain on the amplitude is complicated
and in general not known in closed form, one is forced to approximate.  
This is typically done by characterizing the response of the nonlinear
medium via two main parameters:
the small signal gain, $G\_{ss}$
(namely, the gain experienced by signals of small amplitude)
and the saturated gain, $G\_{sat}$
(namely, the gain experienced by signals of large amplitude).
Of these two parameters,
$G\_{sat}$ can be obtained indirectly, 
because, for each system configuration, it must 
compensate exactly the total losses per cavity round-trip.
The value of $G\_{ss}$, however, is hard to characterize
theoretically or experimentally.
Note that 
the value of $G\_{sat}$ depends on the specific system configuration,
most notably on the pump power.
Note also that gain relaxation is also present, but 
occurs over much slower temporal scales than those of a single optical pulse.
Since we are interested in the behavior of pulses near steady-state,
this difference in the time scales is not a concern.

The gain dynamics is typically approximated by choosing a heuristic 
gain response function that interpolates between the
small signal gain at low powers and the saturated gain at high powers.
%
In particular, it was shown that choosing a Lorentzian-shape function
yields a good approximation for the gain saturation
(e.g., see Ref.~\cite{JOSAB16p1999,JSTQE6p1173}).
Let $E(Z,T)$ be the slowly varying envelope of the electric field,
$Z=z\_{lab}$ is the physical propagation distance and 
$T=t\_{lab}- z\_{lab}/v_g$ is the retarted time
[that is, the time in a reference frame that moves with the
group velocity $v_g=1/k'(\omega)$ of the pulse].
Neglecting all other effects for the moment, 
the combination of linear gain and gain saturation 
yields the following equation for the evolution of the electric field
envelope inside the Ti:sapphire crystal:
\[
\partialderiv EZ= \frac{G\_{ss}}{1+|E/E_o|^2}\,E\,,
\label{e:gain} 
\]
where 
$E_o$ represents some appropriate reference amplitude.
As in~\cite{RQE44p428}, 
we now expand the fraction $1/(1+|E/E_o|^2)$ 
in Taylor series near $E=0$.
Retaining terms up to next-to-leading order,
the right-hand side of \eref{e:gain} is then replaced by
$G\_{ss}\big(1-|E|^2/|E_o|^2)\,E$.

Assuming that the gain spectrum does not change appreciably with
the pump power over the operating range of the laser
(which is a good approximation in Ti:sapphire lasers),
one can neglect the second derivative of the cubic term when
reinstating the time dependence.  
Then, recalling that the gain dynamics is confined to the crystal
and combining the gain dynamics with the dispersive and nonlinear effects 
and the loss arising of the mirrors
yields the following CGLE in dimensional form
with distance-dependent coefficients:
\[
i\partialderiv EZ -\frac12 (k''(Z) -i\beta(Z))\partialderiv[2]ET
  + \big( \gamma(Z)
  + i G\_{nl}(Z) \big) |E|^2E  = iG_o(Z)E\,,
\label{e:NLS+DM+gain}
\]
where $k''(Z)$ and $\gamma(Z)$ are respectively 
the dispersion coefficient and the nonlinear coefficient,
while $\beta(Z)$ quantifies band-limited gain.

The coefficients in~\eref{e:NLS+DM+gain} are conveneniently 
parametrized by introducing the ``indicator'' function $I(Z)$, 
which equals~1 for values of $Z$ corresponding to locations inside the 
Ti:sapphire crystal and 0~otherwise. 
The coefficients in~\eref{e:NLS+DM+gain} can then be written as:
\bse
\label{e:DMcoeffs}
\bea
k''(Z)= k''\_{Ti:s}\,I(Z) + k''\_{dcm}\,(1-I(Z))\,,
\label{e:DMk''}
\\
\beta(Z)= \beta\_{Ti:s}\,I(Z)\,,\qquad
\gamma(Z)= \gamma\_{Ti:s}\,I(Z)\,,
\\
G_o(Z)= G\_{ss}\,I(Z) + G\_{dcm}\,(1-I(Z))\,,\qquad
G\_{nl}(Z)= \frac{G\_{ss}-G\_{sat}}{|E_o|^2}\,I(Z)\,,
\nonumber\\[-1ex]
\label{e:DMgain}
\eea
\ese
where the constants $k''\_{Ti:s},\gamma\_{Ti:s}>0$ are
the dispersion and nonlinear coefficient of the Ti:sapphire crystal;
$k''\_{dcm}<0$ quantifies the dispersion of the mirrors, 
prisms and output coupler, as well as that of the air
(which is experienced during the free-space propagation of the pulse
in the cavity), and $G\_{dcm}<0$ does the same for the loss.
Note that, as written, \eref{e:DMk''} and~\eref{e:DMgain}
model the dispersion compensation and linear loss as taking place
uniformly over the whole length of the cavity not occupied by the 
Ti:sapphire crystal.
This is obviously not the case.   
Nonetheless, for the purposes of the averaging it is immaterial 
whether these effects occur at discrete locations or whether
they are distributed, because all the processes occurring 
outside the crystal are linear.

\paragraph{Nondimensionalization and the DMGLE.}

Next we proceed to nondimensionalize~\eref{e:NLS+DM+gain}.
This requires more care than is necessary for the 
NLSE or the CGLE,
because the dispersive effects are distance-dependent.
It also requires more care than is necessary for the DMNLSE,
because the nonlinear effects are also distance-dependent here.

Let $x= (t\_{lab}- z\_{lab}/v_g)/t_*\,$ and $t= z\_{lab}/z\_*$
be respectively the dimensionless retarded time
and the dimensionless propagation distance,
where $z_*$ is a typical distance (defined below) and
$t_*$ is a typical time scale (e.g., the pulse duration).
Also, let $q= E/E_*$ be the dimensionless 
slowly-varying complex envelope of the electric field,
where $P_*=|E_*|^2$ is a typical pulse power.
Taking into account the fact that nonlinearity only acts 
inside the crystal, we then set
$z_*= (L\_{Ti:s}/L_a)\,z\_{NL}$, where
$L\_{Ti:s}$ is the length of the Ti:sapphire crystal,
$L_a$ is the total length of the cavity
and $z\_{NL}= 1/(\gamma\_{Ti:s}P_*)$ is the distance
at which nonlinear effects become relevant.
These changes of dependent and independent variables,
transform~\eref{e:NLS+DM+gain} into 
the one-dimensional version of~\eref{e:CGLE+DM}, 
with
\bse
\bea
p(\zeta)= d(\zeta) - ib(\zeta)\,,\quad
d(\zeta)= -k''(\zeta)/k''_*\,,\quad
b(\zeta)= \=b\,I(\zeta)\,,
\\
n(\zeta)= (1 + i \=c)\,I(\zeta)/\=I\,,\quad
g(\zeta)= G_o(\zeta)\,z_*\,
\eea
\ese
(with a slight abuse of notation),
where $\zeta= t/t_a$ as before,
and where $k''_*= t_*^2/z_*$, $\=b=\beta\_{Ti:s}\=I/k_*''$ and
$\=c= (G\_{ss}-G\_{sat})/(\gamma\_{Ti:s}|E_o|^2)$.
[Note that $t_a= L_a/z\_*$ and $\=I=L\_{Ti:s}/L_a$.]
Without repeating the perturbation expansion of section~\ref{s:dmgle},
we can then apply its results to conclude that, to leading order, 
the dimensionless slowly varying envelope of the electric field
is given by
\bea
q(x,t)= \F^{-1}\big[\^u(k,t)\,\e^{i\Xi(t/t_a)\,k^2/2}\big]\,
\\[-1ex]
\noalign{\noindent with}
\Xi(\zeta) = \int_0^\zeta\big[d(\zeta')-\=d-i\big(b(\zeta')-\=b\big)\big]\,\d\zeta'\,,
\eea
and where the slowly-varying core 
$u(x,t)= \F^{-1}[\^u(k,t)]$ solves the following DMGLE:
\bea
i\partialderiv ut + \frac12 (\=d-i\=b)\partialderiv[2]ux
\nonumber
\\
\kern4em{ }
   + (1+i\=c)\iint u_{(x+x')}u_{(x+x'')}u^*_{(x+x'+x'')}R_{(x'x'')}\,\d x'\d x''
   = i\=g\,u\,,
\label{e:DMGLE}
\eea
with $\=b,\=c,\=d,\=g\in\Real$, and
\bse
\bea
R(x'x'')= \iint \e^{-ik'x'-ik''x''}r(k'k'')\,\d k'\d k''\,,
\\
\noalign{\noindent with}
r(\xi)= \frac1{(2\pi)^2\=I}\int_0^1\,\e^{i\Xi(\zeta)\xi}I(\zeta)\,\d\zeta\,.
\eea
\ese
If $p(\zeta)\in\Real$ [i.e., $b(\zeta)=0$], 
it is $r^*(\xi)=r(-\xi)$ and $R^*(\xi)= R(-\xi)$.
When $\=b=\=c=\=g=0$, \eref{e:DMGLE} reduces to the DMNLSE
derived in \cite{OL23p1668,OL29p1808}.

The multiple-scale expansion is of course
formally justified only when the expansion parameter $t_a$ is small.
Nonetheless, 
the DMNLSE --- obtained under the same assumptions by neglecting
the gain and loss dynamics --- has been shown to provide a good qualitative 
(and in some cases even quantitative) description of the actual behavior 
of pulses in Ti:sapphire femtosecond lasers even for system configurations 
where $t_a=O(1)$ \cite{PRL94p243904}.  
One therefore expects that the DMGLE will also be a good 
model even in those situations.
 
As in section~\ref{s:dmgle}, the coefficients 
$\=b$, $\=c$, $\=d$ and $\=g$ could also depend on 
the evolution variable~$t$.
Such situations occur when the average dispersion, 
gain and gain saturation exhibit variations over slower temporal scales 
compared to the characteristic duration of the pulses in the cavity.
Note also that 
the average dispersion depends on the details of the
Kerr-lens process in the crystal, on the mirrors, 
the prisms and on the total cavity length.
If $\=d={}$const, one can make its value unity 
by choosing $t_*^2=\=k''z_*$. 
However, the general form is convenient if one wants to
compare systems having different values of 
average dispersion (as in Ref.~\cite{PRL94p243904}), 
since in this case one can do so within 
the framework of a single DGMLE, without having to go back to the
CGLE and choose different normalizations for each case.

\paragraph{Two-step maps.}

For the piecewise-constant two-step maps defined by~\eref{e:DMcoeffs}, 
the kernels $r(\xi)$ and $R(\xi)$ in~\eref{e:DMGLE} can be computed explicitly,
and assume a very simple form.
Taking the origin of the map at the output coupler,
we have
\[
I(\zeta)= \cases{0 &$\zeta\in[0,\theta/2)\cup[1{-}\theta/2,1)$\,,\\
  1 &$\zeta\in[\theta/2,1{-}\theta/2)$\,,}
\]
where $0<\theta<1$ represents the fraction of the cavity length 
not occupied by the Ti:sapphire crystal.
This yields $\=I= 1-\theta$.
Recalling that the linear effects acting on the pulse can be distributed 
throughout the portion of the cavity not occupied by the crystal,
we then parametrize the zero-mean part of the dispersion as
\cite{OL23p1668,PRL94p243904}
\[
\Delta d(\zeta)= \cases{s/2\theta 
    &$\zeta\in[0,\theta/2)\cup[1{-}\theta/2,1)$\,,\\
  -s/[2(1-\theta)] &$\zeta\in[\theta/2,1{-}\theta/2)$\,,}
\]
where the map strength parameter~$s$
quantifies the $L_1$-norm of $\Delta d(\zeta)$,
as discussed in section~\ref{s:symmetries}.\,
(This definition differs from the previous one 
\cite{OL23p1668,PRL94p243904} by a factor~4.)
Then
\[
\Xi(\zeta)= \cases{
  \Xi_+\,\zeta &$\zeta\in[0,\theta/2)$\,,\\
  \Xi_-\,(\zeta-1/2) &$\zeta\in[\theta/2,1{-}\theta/2)$\,,\\
  \Xi_+\,(\zeta-1) &$\zeta\in[1{-}\theta/2,1)$\,,}
\]
where
$\Xi_+=2(s+ib_o)/\theta$ and
$\Xi_-=-2(s+ib_o)/(1-\theta)$,
and with
$b_o= \=b\theta t_a/2$.
After some straightforward calculations, one then obtains the kernels as
\bse
\bea
r(\xi)= \frac1{(2\pi)^2}\,\sinc[(s+2ib_o)\xi/4]\,,
\label{e:kernel2step}
\\
R(\xi)= \frac2{\pi(s+ib_o)}\ci[4\xi/(s+ib_o)]\,,
\eea
\ese
where $\sinc(y)= (\sin y)/y$ and 
$\ci(y)= \int\nolimits_y^\infty \cos\,t\,\d t/t$
is the cosine integral.
The functional form of both kernels is identical to that of the DMNLSE. 
(The kernel in Ref.~\cite{PRL94p243904} differs 
by~\eref{e:kernel2step} by an overall
multiplicative constant due to the different choice of normalizations.)
In the DMGLE, however, the kernels $r(\xi)$ and $R(\xi)$ appear multiplied 
by the complex factor $1{+}i\=c$, which has important consequences
on the equation (as we show in section~\ref{s:symmetries}) and
on its solutions (as we show in section~\ref{s:solutions}).
Note that when $\=b=0$ both kernels are real, as for the DMNLSE without 
linear gain and loss~\cite{OL23p1668},
and the specific details 
of the functions $\Delta d(t)$ and $I(t)$ do not affect 
the DMGLE nor its kernels.
For this class of systems the DMGLE is a universal equation, 
in that it provides a unified description for the dynamics
of different systems (like the NLSE, CGLE and DMNLSE).

\section{Symmetries, rate equations and linear modes}
\label{s:symmetries}

We now look at the properties of the DMGLE.
For simplicity, in the remainder of this work we will restrict 
our attention to the form of the DMGLE that appears in femtosecond lasers,
and to the case in which
the coefficients $\=b$, $\=c$, $\=d$ and $\=g$ appearing 
in~\eref{e:DMGLE}
are independent of time.
We emphasize however that many of the results that follow can be extended 
in a straightforward way to the more general form of the 
DMGLE~\eref{e:DMGLEo}.

\paragraph{Symmetries.}

The DMGLE~\eref{e:DMGLE} enjoys a number of symmetries.
Let $u(x,t)$ be any solution of~\eref{e:DMGLE}.
Then $u_\epsilon(x,t)$ is also a solution of~\eref{e:DMGLE} 
$\forall\epsilon\in\Real$, where: 
\begingroup
\advance\leftmargini 0.2em
\begin{enumerate}
\item
Phase invariance:
$u_\epsilon(x,t)= \e^{i\epsilon}u(x,t)$.
(The same symmetry applies for the NLSE, CGLE and DMNLSE.)
\item
Space translations:
$u_\epsilon(x,t)= u(x-\epsilon,t)$.
(The same symmetry applies for the NLSE, CGLE and DMNLSE.)
\item
Time translations. 
$u_\epsilon(x,t)= u(x,t-\epsilon)$.
(The same symmetry applies for the DMNLSE.  
It also applies for the NLSE and CGLE in the case of constant coefficients.
In general~\eref{e:CGLE+DM} is \textit{not} invariant 
under time translations, however.)
\item
``Chirp'': 
$\^u_\epsilon(k,t)= \e^{i\epsilon k^2/2}\^u(k,t)$
is a solution of the equation with $r(\xi)$ replaced by
$r_\epsilon(\xi)=\e^{i\epsilon\xi}r(\xi)$.
(The name comes from the fact that quadratic variations of the pulse phase
are referred to as ``chirp'' in optics~\cite{Agrawal,MollenauerGordon}.)
This symmetry, which is the same as that for the DMNLSE~\cite{OL23p1668}, 
has no analogue for either the NLSE or the CGLE.
\item
Galilean boosts
[generalized Galilean boosts if $\=d(\cdot)$ is time-dependent]:
if $\=b=0$,
\[
u_\epsilon(x,t)= \e^{i[\epsilon tx - \half\epsilon^2m(t)]}
  u(x-\epsilon m(t),t)
\label{e:Galilean}
\]
with $m(t)= \int\nolimits_0^t \=d(t')\,\d t'$.
(The NLSE and CGLE with varying coefficients admit a similar
invariance~\cite{JOSAB16p1628}.
For the DMNLSE, as well as for the NLSE and CGLE with constant coefficients,
it is simply $m(t)= \=d\,t$.
For the DMGLE with $\=b\ne0$, \eref{e:Galilean} applies only for those
solutions that can be analytically continued off the real $x$-axis.)
\item
Generalized scaling symmetry.
If $u(x,t;\=g,s)$ is a solution of~\eref{e:DMGLEo}, 
so is
\[
u_\epsilon(x,t;\=g,s)= a\,u(ax,a^2t;\=g/a^2,a^2s)\,
\label{e:scaling}
\]
with $a=1+\epsilon$
and where $s$ is the map strength parameter, as discussed below.
(The parametrization of~\eref{e:scaling}
is chosen so that the value $\epsilon=0$ reproduces the original solution,
as in all other cases.)
\end{enumerate}
\endgroup
These invariances will allow us in section~\ref{s:solutions} 
to construct a two- or three-parameter family of solutions 
of the DMGLE (depending on whether $\=b\ne0$ or $\=b=0$, respectively)
from a single stationary solution,
in a similar way as for the NLSE, DMNLSE and CGLE.
As we show later, however, 
unlike for the NLSE and DMNLSE (and like for the CGLE)
the scaling invariance does \textit{not} generate a one-parameter
family of solutions.
Note also that, 
as a result of the chirp symmetry, the constant $\Xi_o$ in~\eref{e:mdef}
can be chosen arbitrarily, and does not affect the solution $q(x,t)$
of the original problem~\eref{e:CGLE+DM}, since it simply amounts to a 
redefinition of $\^u(k,t)$.
(Proper choice of $\Xi_o$ can be useful to make the kernels simpler,
however.)

\paragraph{Map strength.}

In the statement of the scaling invariance above we used the obvious 
fact that the solutions of the DMGLE~\eref{e:DMGLEo} depend 
parametrically on~$\=g$.
One should realize, however, that, like for the DMNLSE, 
solutions of the DMGLE
also depend on a parameter~\,$s$ called the \textit{map strength},
which quantifies the size of the zero-mean dispersion variations.
The map strength
(which for two-step maps was introduced in section~\ref{s:lasers})
can be formally defined for any map
as the $L_1$-norm of $\Delta d$
(in the DMGLE as for the DMNLSE~\cite{PRA75p53818}),
namely:
\begin{equation}
s= \overline{|\Delta d|}=
  \int_0^1|\Delta d(\zeta)|\,\d\zeta\,.
\label{e:sdef}
\end{equation}
One can then obtain explicitly the dependence of the kernels 
$r(\xi,t)$ and $R(x',x'',t)$ on~$s$ 
by writing $\Delta d(\zeta)$ 
[and consequently $\Xi(\zeta)$ via~\eref{e:mdef}]
in terms of normalized functions.
That is, given any choice of map $\Delta d_o(\zeta)$, 
we can define the normalized function
$\Delta d\_{ref}(\zeta)= \Delta d_o(\zeta)/\overline{|\Delta d_o|}$.
One can then introduce a one-parameter family of dispersion functions
\bea
\Delta d_s(\zeta)= s\,\Delta d\_{ref}(\zeta)\,.
\eea
In this way one can study the behavior of solutions for 
different values of map strengths within the framework of the DMGLE,
without needing to go back to~\eref{e:CGLE+DM}. 
When necessary, 
we express this dependence explicitly by writing the solution 
of~\eref{e:DMGLEo} as $u(x,t;\=g,s)$, 
like we did in~\eref{e:scaling}.
Of course, in the limit $s\to0$ 
the DMGLE~\eref{e:DMGLE} 
reduces to the CGLE with constant coefficients.
Using the map strength parameter, it is now easy to show that, 
if $\=d,\=g$ and $K$ are independent of~$t$,
the generalized scaling invariance~\eref{e:scaling} holds.
Of course this invariance reduces to those of the DMNLSE, CGLE and NLSE 
in the cases $s=0$ and/or $\=b=\=c=\=g=0$.

\paragraph{Rate equations.}

Like the CGLE, and unlike the NLSE and DMNLSE, 
the DMGLE is not a Hamiltonian system.
Hence it is not possible to use Noether's theorem to derive
conservation laws from the symmetries of the equation
as done for the NLSE~\cite{Kath1997}. 
Nonetheless, by analogy with the NLSE/DMNLSE, it is still 
possible to associate each symmetry with a rate equation.
For brevity we only list the first three such equations,
and only in the special case $\=b=0$:
\bse
\label{e:rate}
\bea
\fl
\deriv{ }t\int |u|^2 \d x \, = \,
  2\=g\int|u|^2 \d x
    - 2\=c\iiint u^*_{(x)}u_{(x+x')}u_{(x+x'')}u^*_{(x+x'+x'')}
        R_{(x'x'')} \d x'\d x''\d x\,,
\label{e:energy}
\\
\fl
\deriv{ }t \Im \int u^*\partial_x u\,\,\d x \, = \,
  2\=g\,\Im \int u^*\partial_x u\,\,\d x
   - 2\=c\,\Im \int \big(\partial_x u^*\big)\,
      u_{(x+x'')}u_{(x+x'')}u^*_{(x+x'+x'')}
       R_{(x'x'')} \d x'\d x''\d x\,,
\nonumber
\\[-1ex]
\label{e:momentum}
\\
\fl
\frac12 \deriv{ }t \big[ \=d\int|\partial_x u|^2 \d x
   - \iiint u^*_{(x)}u_{(x+x')}u_{(x+x'')}u^*_{(x+x'+x'')}
     R_{(x'x'')} \d x'\d x''\d x
  \big]=
\nonumber
\\
  \=g\,\big[ \=d\int\big|\partial_x u\big|^2 \d x  
  - 2\int u^*_{(x)}u_{(x+x')}u_{(x+x'')}u^*_{(x+x'+x'')}
        R_{(x'x'')} \d x'\d x''\d x \,\big]
\nonumber
\\\kern3em{ }
   + \=c\,\big[ \,
        \=d\,\Re\iiint\big(\partial_x^2u^*\big) 
          u_{(x+x')}u_{(x+x'')}u^*_{(x+x'+x'')} R_{(x'x'')} \d x'\d x''\d x
\nonumber
\\\kern6em{ }
      + 2\int\big| 
	\iint u_{(x+x')}u_{(x+x'')}u_{(x+x'+x'')}^*R_{(x'x'')}\d x'\d x''
      \big|^2\d x\,
      \big]\,,
\label{e:hamiltonian}
\eea
\ese
where for brevity we used the notation 
$\partial_x u= \partial u_{(x)}/\partial x$
etc.
Each of the above rate equations reduces to those of the NLSE
\cite{Kath1997,Sulem&Sulem}, 
CGLE and DMNLSE in the appropriate limits, of course.
(It appears however that, even in the simpler version obtained for $s=0$, 
that is for the CGLE,
the third rate equation is not widely known; cf.\ Ref.~\cite{PRL79p4047}.)\,\ 
In particular, for the NLSE and DMNLSE, the right-hand-side is zero,
and the integrals on the left-hand side of equations~\eref{e:rate},
which are then the conserved quantities of the corresponding equation,
are respectively the pulse energy, the momentum and the Hamiltonian.
Because of the physical meaning associated with these integrals,
the rate equations for the NLSE and DMNLSE have been extensively used 
to study the evolution of various pulse characteristics.
The rate equations of the DMGLE should therefore prove to be
similarly useful.

\paragraph{Linear modes.}

If $u(x,t)$ is any solution of~\eref{e:DMGLE} and 
$u_\epsilon(x,t)= u(x,t)+ \epsilon w(x,t)$ is also a solution, then
$w(x,t)$ belongs to the nullspace of the linearized DMGLE operator
about $u(x,t)$.  That is, solves the linearized DMGLE $L[w,u]=0$, where
(for $\=b=0$)
\bea
\fl
L[w,u]= i \partialderiv wt + \frac12 \=d \partialderiv[2]wx  - i\=g\,w
 + 2(1+i\=c)\iint
  w_{(x+x')}u_{(x+x'')}u^*_{(x+x'+x'')}R_{(x'x'')}\,
     \d x' \d x'' 
\nonumber
\\\kern9.4em{ }
 + (1+i\=c)\iint
  u_{(x+x')}u_{(x+x'')}w^*_{(x+x'+x'')}R_{(x'x'')}\,
     \d x' \d x''\,.
\nonumber
\\
\label{e:LDMGLE}
\eea
It then follows that,
for each continuous invariance of the DMGLE, there exists a solution 
of the linearized DMGLE in the form
\[
w(x,t)= \partialderiv{u_\epsilon(x,t)}\epsilon\bigg|_{\epsilon=0}\,,
\]
similarly to the CGLE, NLSE and DMNLSE~\cite{PRA75p53818}.
Then, by simply applying the invariances listed above, 
we obtain the following set of linear modes
and generalized linear modes:
\bse
\bea
w_1= iu\,,\qquad
w_2= -\partialderiv ux\,,\qquad
w_3= -\partialderiv ut\,,\qquad
w_4= ix u\,,
\\
w_5= u + x\partialderiv ux + 2s\partialderiv us -2\=g \partialderiv u{\=g}\,,
\eea
\ese
corresponding respectively to phase rotations, space translations,
Galilean boosts and scaling transformations as given above.
(Of course $w_4$ only applies when $\=b=0$.  
Note also that the chirp symmetry requires changing the kernels,
so the above analysis does not apply.)
The above linear modes and generalized modes satisfy the relations
\bse
\bea
L[w_1,u]= L[w_2,u]= L[w_3,u]= 0\,,\\ 
L[w_4,u]= iw_2\,,\qquad
L[w_5,u]= 2iw_3\,.
\eea
\ese
In the special case of traveling wave solutions, 
time translations are simply a composition of space translations and
phase rotations.
As a result, for traveling wave solutions $w_3$ can be expressed as
a linear combination of $w_1$ and $w_2$, and is not an independent mode.
This is consistent with the familiar result \cite{Sulem&Sulem}
that for the NLSE in $N$ spatial dimensions the linearized operator 
around a solitary wave solution has a zero eigenvalue of multiplicity 
$2N{+}2$.
(The result also applies to the DMNLSE~\cite{PRE62p4283}.)

The linearized DMGLE is of course useful to study the stability
of solutions.  
As in the NLSE and DMNLSE, the linear modes of the DMGLE 
generate changes in the solution parameters, 
and can be used to quantify the change of these parameters
under perturbations~\cite{Iannone,PRA75p53818}. 
Because of the presence of a nontrivial nullspace, 
secularities arise even if all eigenvalues of $L$ 
have zero or negative real part, and as a result the change in 
the solution of the DMGLE does not in general remain bounded in time.
As usual, secular terms are removed by taking the solution parameters
to be slowly dependent on time and determining their evolution by projecting
the perturbation onto the linear modes.
The existence of a generalized mode corresponding to the 
generalized scaling invariance
is then important because the coupling between amplitude and phase in the 
NLSE/DMNLSE is the mechanism whereby the
variance of noise-induced phase perturbations grows cubically
\cite{OL15p1351} 
(similarly to the well-known coupling between frequency and timing
jitter \cite{OL11p665}).
Even though the relation between amplitude and phase 
is broken in the DMGLE due to the presence of gain and gain saturation,
the existence of a generalized mode associated to the phase
suggests that the noise-induced phase variance could still grow cubically.

\section{Solutions, stability and parameter dependence}
\label{s:solutions}

We now discuss special solutions of the DMGLE.
Again, for concreteness, we restrict our consideration to 
the specific form of the DMGLE arising in femtosecond lasers, 
namely~\eref{e:DMGLE} with constant coefficients.

\paragraph{Soliton solutions of the DMGLE.}

We start by looking for stationary solutions, that is, 
solutions in the form 
\[
u(x,t)= f(x)\,\e^{i\lambda^2 t/2}\,.
\label{e:DMSansatz}
\]
[Recall that a family of traveling wave solutions
can be obtained from~\eref{e:DMSansatz}
by applying the invariances of the DMGLE.]
Substituting the Fourier transform of the above ansatz in~\eref{e:DMGLE}
yields the following nonlinear integral equation
\bea
[\,\half\lambda^2 + \half(\=d -i\=b)k^2 + i\=g\,]\,\^f 
= 
\nonumber
\\
{ }\kern8em (1+i\=c)\iint \^f_{(k+k')}\^f_{(k+k'')}\^f^*_{(k+k'+k'')}r_{(k'k'')}
   \,\d k'\d k''\,.
\label{e:DMSinteqn}
\eea
Although closed-form solutions are not available unless $r(\,\cdot\,)={}$const,
\eref{e:DMSinteqn} can be efficiently integrated numerically
(see~Appendix).
Note also that, similarly to the DMNLSE
\cite{PRA75p53818,OL27p939},
fast numerical methods can be used for the calculation of 
the double integral in Eqs.~\eref{e:DMGLE} and~\eref{e:DMSinteqn}
and thus to solve numerically both the integral equation~\eref{e:DMSinteqn}
and the DMGLE~\eref{e:DMGLE} itself
(again, see Appendix).
Hence, the computational complexity of the DMGLE is no less and no greater
than that of the original, un-averaged equation~\eref{e:CGLE+DM}.

Of course, solutions of~\eref{e:DMSinteqn} 
are not solitons in the mathematical sense of the term\break 
(that is, solutions corresponding to the discrete spectrum 
of the scattering problem associated to the given nonlinear 
partial differential equation~\cite{AblowitzSegur}),
but rather solitary waves.  
As common in physics and optics, however
\cite{Agrawal,Chaos10p515,SIAMREV48p629,MollenauerGordon,YeCundiff},
we will refer to such a pulse as a soliton --- or, in the case of 
the DMNLSE and DMGLE, as a dispersion-managed soliton (DMS).
Importantly, while for the NLSE and the DMNLSE
one can find solutions for any real~$\lambda$,
for the CGLE and the DMGLE nontrivial solutions exist only 
for \textit{discrete} values of~$\lambda$.

\begin{figure}[t!]
\vskip1.4\smallskipamount
\centerline{\includegraphics[width=0.666\textwidth]{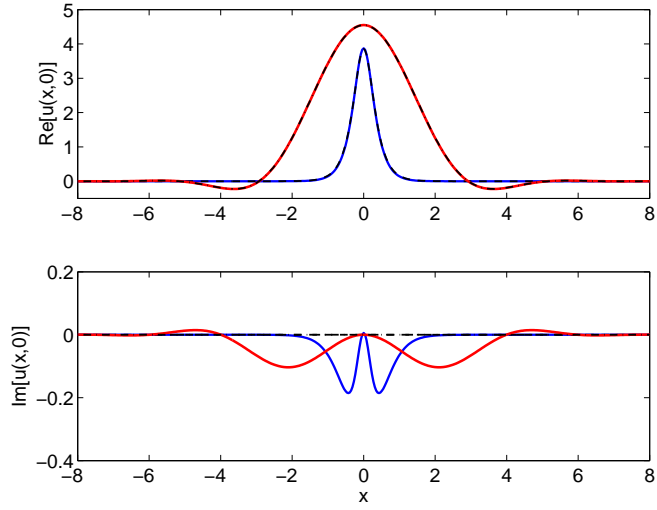}}
\caption{Shape of the solutions of~\protect\eref{e:DMSinteqn}.
\let\==\bar
Solid blue line: CGLE ($s=0$, $\=g=2$, $\=b=\=c=0.2$);
dashed black line: NLSE ($s=\=b=\=c=\=g=0$);
thick solid red line:  DMGLE ($s=4$, $\=g=2$, $\=b=\=c=0.2$);
dot-dashed black line: DMNLSE ($s=4$, $\=b=\=c=\=g=0$).
In all cases $\=d=1$.
The value of $\lambda$ for the NLSE and the DMNLSE was chosen so that 
their solutions have the same peak amplitude as those of the
CGLE and DMGLE, respectively.
The imaginary part of the solutions of both the NLSE and the DMNLSE 
is identically zero, while the real part is almost indistinguishable
from that of solutions of the CGLE and DMGLE, respectively.}
\label{f:dmsreim}
\vskip2\medskipamount
\end{figure}
\begin{figure}[t!]
\centerline{\kern1em\includegraphics[width=0.644\textwidth]{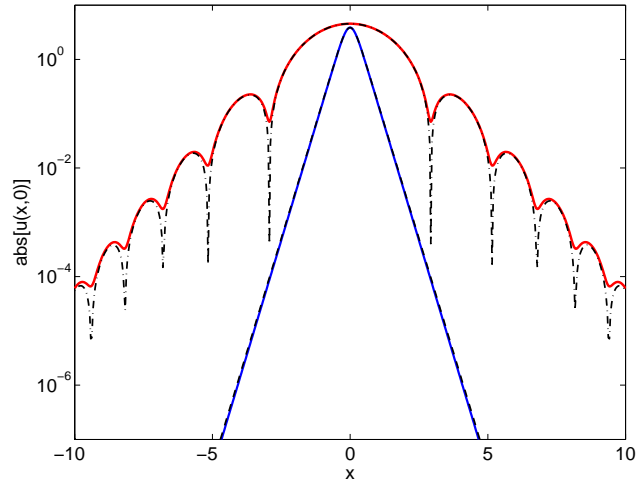}}
\caption{Absolute value of the stationary solutions in semi-logarithmic scale 
for the same cases as in \fref{f:dmsreim}.
(Line colors and styles are also the same.)
Solutions of the DMGLE have exponentially decaying oscillating tails, 
like those of the DMNLSE, 
with the frequency of the oscillations increasing along the tails.
Unlike the solutions of the DMNLSE, however,
solutions of the DMGLE do not appear to posseses zeros,
due to the presence of a non-zero imaginary part.}
\label{f:dmsabs}
\end{figure}

\Fref{f:dmsreim} shows the real and imaginary parts of 
stationary solutions of the NLSE, CGLE, DMNLSE and DMGLE,
while \fref{f:dmsabs} shows the modulus of the same solutions
in semi-logarithmic scale.
The values of $\lambda$ that yield
the unique solution with the given value of the parameters
for the CGLE and the DMGLE are respectively
$\lambda\_{cgle}=3.86$ and $\lambda\_{dmgle}= 4.43$,
as obtained numerically using the methods described in appendix.
For the NLSE and the DMNLSE, the value of $\lambda$ was chosen so that
the corresponding solutions have the same peak amplitude as those of
the CGLE, respectively.
Such values are easily obtained by noting that,
for both the NLSE and DMNLSE, $\lambda=\max_{x\in\Real}|u(x,0)|$.
In all cases, 1024 Fourier modes were used in the simulations.

It should be clear from these figures that the solutions of the DMGLE combine 
some features of the solutions of the CGLE with some of the 
solutions of the DMNLSE.
For example, it is evident from both \fref{f:dmsreim} and \fref{f:dmsabs} 
that the shape of the soliton solutions of the DMGLE is remarkably similar 
to that of the soliton solutions of the DMNLSE.
At the same time, however, \fref{f:dmsreim} shows that, 
like the solutions of the CGLE, and unlike those of the DMNLSE,
a small but nonzero imaginary component appears. 
As a consequence, 
the solutions of the DGMLE do not appear to possess zeros, 
unlike those of the DMNLSE.

\paragraph{Parameter dependence.}

To further explore the similarities and differences between solutions 
of the DMGLE, CGLE and DMNLSE, 
we next look at their parameter dependence.
Even though the parameter $\lambda$ 
appears in identical way in the integral equation
for the solutions of the DMNLSE and DMGLE,
it nonetheless plays a very different role in the two cases.
To see why, we briefly look at the case $s=0$
and compare solutions of the NLSE and the CGLE.
When transformed back to the physical domain,
the solutions of~\eref{e:DMSinteqn} are, in this case, 
\bse
\bea
u\_{nlse}(x,t)= \sqrt{\!\=d}\,a\,\sech(ax)\,\e^{ia^2\=dt/2}\,,
\label{e:NLSsoliton}
\\
u\_{cgle}(x,t)= \sqrt{\!\=d}\,\big[A\,\sech(\eta x)\big]^{1+i\nu}
  \,\e^{i\mu\=dt}\,,
\label{e:CGLEsoliton}
\eea
\ese
where $a$ is an arbitrary real constant, while~\cite{PRE53p1190}
\bea
A= \sqrt{\frac{3\nu(1+\=b^2)\,\=g}{\=c\alpha}}\,,\quad
\eta= \sqrt{\frac{2\=g}\alpha}\,,\quad
\mu= (1-\nu^2-2\nu\=b)\,\frac{\=g}\alpha\,,
\nonumber
\\
\noalign{\noindent with}
\alpha= 2\nu-\=b(1-\nu^2)\,,\quad
\nu=\frac{-3(1-\=b\=c)+\sqrt{9(1-\=b\=c)^2+8(\=b+\=c)^2\strut}}{2(\=b+\=c)}\,.
\label{e:c}
\eea
The functional form of~\eref{e:NLSsoliton} and~\eref{e:CGLEsoliton}
is similar.
The key difference, however, is that \eref{e:NLSsoliton} represents
a one-parameter family of solutions, since $a$ 
[the soliton amplitude, which is inversely proportional to the pulse width
and coincides with the soliton eigenvalue in~\eref{e:DMSinteqn} as well as
with the scaling parameter in~\eref{e:scaling}]
can take any real value.
In contrast, all parameters in~\eref{e:CGLEsoliton} are uniquely
determined by the coefficients in the CGLE.  
Thus, none of them can play the role of the scaling parameter
in~\eref{e:scaling}.
Instead, that role for~\eref{e:CGLEsoliton} is played by $\sqrt{\=g}$. 

A similar situation arises for the DMNLSE and the DMGLE.
Namely, for the DMNLSE the eigenvalue~$\lambda$ 
is also the soliton amplitude and the scaling parameter,
and is in one-to-one correspondence with the pulse energy
(even though the relation is not simply
of direct proportionality as for the NLSE~\cite{PRA75p53818}).
For the DMGLE, in contrast, the value of $\lambda$ 
is completely determined by that of the coefficients 
in the DMGLE.
Thus, in both the CGLE and the DMGLE, the addition of gain and loss
dynamics seems to simply pick out one particular solution in the
family of solutions of the corresponding conservative model
(NLSE and DMNLSE, respectively),
without significantly altering its form --- at least for moderate values
of the gain/loss parameters.
 
\begin{figure}[t!]
\vskip\smallskipamount
\centerline{\includegraphics[width=0.644\textwidth]{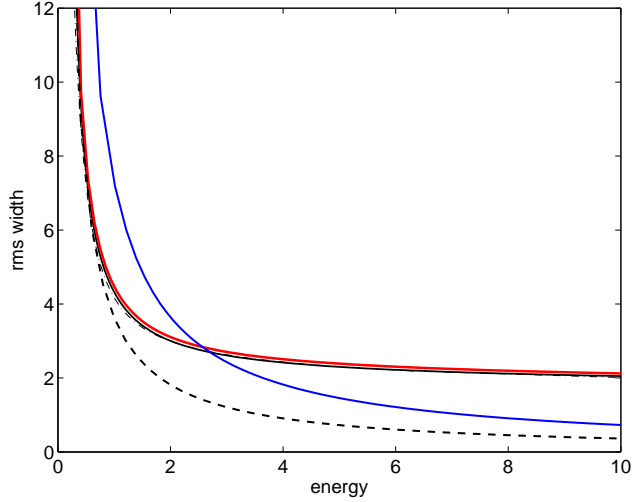}}
\caption{\let\==\bar
Root-mean-square (rms) width of the stationary solutions versus
their energy for the NLSE, CGLE, DMNLSE and DMGLE
(line colors and styles are as in figures~\ref{f:dmsreim} and~\ref{f:dmsabs}).
The data for the NLSE and the CGLE are obtained respectively
from the first and the second of~\eref{e:Evstau},
while those for the DMNLSE and the DMGLE are obtained from numerical
solution of the integral equation~\eref{e:DMSinteqn} 
for different values of $\lambda$ (DMNLSE) and $\=g$ (DMGLE).
The value of all the other parameters is the same as in figures~\ref{f:dmsreim}
and~\ref{f:dmsabs}.}
\label{f:energywidth}
\end{figure}

The above observations are corroborated by looking at the relation
between the energy of the pulse and its root-mean-square width, 
defined respectively as
\bea
E= \int|u|^2\,\d x\,,
\qquad
\tau= \sqrt{M_2/E}\,,
\\
\noalign{\noindent where}
M_2= \int x^2|u|^2\,\d x\,.
\nonumber
\eea
For the solutions~\eref{e:NLSsoliton} and~\eref{e:CGLEsoliton} 
of the NLSE and the CGLE, these integral quantities take on 
the following values:
\bse
\bea
E\_{nlse}= 2\=da\,,\qquad 
\tau\_{nlse}= \pi/(\sqrt3\,a)\,,
\label{e:EtauNLSE}
\\
E\_{cgle}= 3\=d\sqrt{\=g\eta}\,\big/{\=c}\,,\qquad 
\tau\_{cgle}= \pi\sqrt{\eta/3{\=g}}\,,
\label{e:EtauCGLE}
\eea
\ese
with $\eta$ uniquely determined from $\=b$ and $\=c$ via~\eref{e:c} 
as before.
Eliminating $a$ from~\eref{e:EtauNLSE}
and $\=g$ from~\eref{e:EtauCGLE} we then obtain
the well-known relations between amplitude and width,
respectively for the NLSE and the CGLE:
\[
E\_{nlse}\tau\_{nlse}= 2\pi\=d/\sqrt3\,,
\qquad
E\_{cgle}\tau\_{cgle}= \sqrt3 \pi\=d\eta/\=c\,.
\label{e:Evstau}
\]
Equivalent relations were derived in~\cite{BC@NLP2002} for the DMNLSE 
using a variational approximation with a Gaussian ansatz:
\bse
\label{e:EtauDMNLSE}
\bea
E\_{dmnlse}= \=d\sqrt{\frac\pi{2a}}\,
  \frac{\sqrt{1+y^2}}{\big\{
    1-\sqrt{1+y^2}\ln\big[(1+\sqrt{1+y^2})/y\big]/2\big\}\,y}\,,
\\
\tau\_{dmnlse}= \sqrt{2a}\,,
\eea
\ese
where $y=a/s$.
Eliminating~$a$ from~\eref{e:EtauDMNLSE} then yields the equivalent 
of~\eref{e:Evstau}.
To the best of our knowledge, however, no variational approach 
has been developed for the CGLE that can be extended to the DMGLE.
In this case, therefore, the relation between energy and width 
must be obtained 
by numerically solving the integral equation~\eref{e:DMSinteqn}.

\Fref{f:energywidth} shows the rms pulse width versus its energy 
for the NLSE, the CGLE, the DMNLSE and the DMGLE.
We see that, as a result of the dispersion management, the 
effects of gain dynamics are reduced compared to the constant dispersion case,
and the similarity between the parameter dependence of
the solutions of the DMNLSE and the DMGLE is even closer than
that among the solutions of the NLSE and CGLE.
[Note also that, for the DMNLSE, the data from~\eref{e:EtauDMNLSE}
is almost indistinguishable from those coming from numerical
solutions of~\eref{e:DMSinteqn}.]
Note however that the curves for the NLSE/DMNLSE are obtained by
varying~$\lambda$, while those for the CGLE/DMGLE by varying~$\=g$.
Indeed, it is crucial to realize that each point in the curves 
for the CGLE/DMGLE is a solution of a \textit{different} 
partial differential equation.
As a result the above comparison may appear to be somewhat artificial
at first.
In practice it is not, however.
This is because solutions with different amplitude 
are experimentally produced by varying the pump power
\cite{PRL94p243904}, which has precisely the effect of
changing the value of the linear gain coefficient --- and thus of $\=g$
--- in the DMGLE~\eref{e:DMGLE}.

\paragraph{Stability.}

In the absence of gain dynamics, 
solutions of the NLSE and one branch of solutions of the DMNLSE 
are stable under perturbations.
In the presence of linear, band-limited gain and gain saturation, 
one would expect the pulse solutions of the CGLE/DMGLE to be 
more stable than those of the NLSE/DMNLSE.
Indeed, solutions of the CGLE with constant coefficients are 
stable for $\alpha>0$ \cite{PRE53p1190}.
(Such is the case for all the solutions in 
figures~\ref{f:dmsreim}--\ref{f:energywidth}.)
We expect that in these situations, amplitude perturbations
will be damped in the DMGLE like they are in the CGLE,
making the stationary pulses of the DMGLE stable.
But of course this conjecture must be verified via careful
analysis.
One way to do so would be to perform a linearized stability analysis,
that is, to look for the eigenvalues of the linearized DMGLE operator
$L[w,u]$ in~\eref{e:LDMGLE} when $u$ is the solitary-wave 
solution~\eref{e:DMSansatz} obtained via~\eref{e:DMSinteqn},
along the lines of what was done for the DMNLSE in 
Refs.~\cite{OL28p1754,Chaos10p539,PRE62p4283,OC246p393}.

\section{Conclusions}
\label{s:conclusions}

In summary, we have derived the dispersion-managed Ginzburg-Landau
equation (DMGLE) as the equation that governs the long-term dynamics 
of systems described by the CGLE with time-dependent coefficients.
In particular we have shown how the DMGLE arises in
Ti:sapphire femtosecond lasers, we discussed the properties
of the equation and of its solutions.
Since the DMGLE is (like the CGLE, NLSE and DMNLSE) a universal equation,
however,
we believe that it will prove to be a unified model for the description 
many other experimental realizations of femtosecond lasers in addition 
to Ti:sapphire, 
such as similariton lasers~\cite{PRL92p213902},
lasers using all-normal dispersion fiber~\cite{OE14p10095},
and those using waveguide arrays~\cite{OL30p2013}.

From a mathematical point of view, perhaps the most important feature of
the DMGLE is that it is amenable to analytical treatment.
In fact, the results of this work open up many interesting theoretical 
questions, as well as some important practical ones.  
In particular, the following are fairly natural open questions:
\begin{itemize}
\item 
Whether a proof of existence of stationary solutions 
is possible (as was done in Ref.~\cite{PRE62p7358} for the DMNLSE).
\item 
Obtaining the asymptotic behavior of the stationary solution as $x\to\pm\infty$
(as was done in Refs.~\cite{OL25p1144,OL26p1535} for the DMNLSE).
\item 
Whether there exist various branches of solutions as a function of $\=d$
as in the DMNLSE (e.g., see Ref.~\cite{Chaos10p515} and references therein).
\item 
Whether there exist multi-pulse solutions 
(as in the CGLE~\cite{AkhmedievAnkiewicz} and the higher-order DMNLSE~\cite{JOSAB19p2876}).
\item 
The stability of all of these solutions
(e.g., see Refs.~\cite{PRE62p4283,OC246p393} and references therein
for the DMNLSE and~\cite{PRE53p1190,JOSAB19p740} for the CGLE).
\item 
Whether there exist ``true'' dispersion-managed solitons
corresponding to the stationary solutions of the DMGLE~\eref{e:DMGLE}; 
that is, whether there exist time-periodic solutions in the original,
unaveraged system~\eref{e:CGLE+DM}.
(Recall that for NLS+DM, radiative losses beyond all orders \cite{PhysD149p80}.)
\item 
Even if true dispersion-managed solitons did not exist in the original,
un-averaged system, it should still be possible to prove that 
solutions of~\eref{e:CGLE+DM} remain $O(t_a)$ close to 
those of~\eref{e:DMGLE} up to times $O(1/t_a)$
(as was done for the DMNLSE~\cite{PHYD152p794}).
\item 
Characterizing the large-$s$ limit of the equation and of its solutions
(as was done in Ref.~\cite{OL26p459} for the DMNLSE).
\item 
Formulating ``slow'' ordinary differential equations 
for the long-term evolution of the pulse parameters 
which bypass the fast dynamics inside each map
(as was done in Ref.~\cite{OL26p1761,PRE62p4283} for the DMNLSE).
\item 
Characterizing pulse collisions and interactions. 
Because neither the CGLE nor the DMNLSE~\cite{JOSAB18p577,PLA260p68}
are integrable, one would expect that the DMGLE is not either,
implying that pulse interactions will be inelastic.
\item 
More generally, since the CGLE with constant coefficients describes 
a remarkably rich variety of physical phenomena, including 
chaos~\cite{PRL95p024101},
it will be interesting to use the DMGLE to see how these phenomena 
are affected by the presence of dispersion and nonlinearity management.
\end{itemize}
Since the Hamiltonian formalism is lost, however,
settling these issues might be significantly more complicated than
for conservative systems such as the NLSE and DMNLSE. 

A more complicated equation of DMNLS type with gain and loss 
was also recently studied in \cite{MJA2008} as a model 
for Ti:sapphire lasers.
Also, a review of different mathematical approaches for the study of 
dispersion management in optical fibers can be found in 
Ref.~\cite{Chaos10p515}.  
All of the works cited in Ref.~\cite{Chaos10p515}, however, deal with 
conservative systems.
To the best of our knowledge,
Ref.~\cite{MJA2008} and the present work are the first to generalize
those tools in order to study dispersion-managed systems with 
significant gain dynamics.

With regard to more practical issues pertaining to 
Ti:sapphire femtosecond lasers, 
we\break first note that
a multi-dimensional version of the DMGLE such as the one
presented in section~\ref{s:dmgle}
could be useful in order to take into
account the transverse dynamics of the pulses in the cavity.
In any case, the DMGLE can now be used to study the sensitivity of pulses
with respect to perturbations, and especially quantum noise, 
using the linear modes and their adjoints to guide
importance-sampled Monte Carlo simulations
along the line of Refs.~\cite{PRA75p53818,OL28p105,SJAM67p1418,SIREV50p523}.
(For an introduction to importance sampling tailored to the study 
of noise in lightwave systems See Ref.~\cite{SIREV50p523}.) 
Such a study will be instrumental to determine the true stability 
properties of pulses in these lasers and consequently to obtain 
the comb linewidth, 
which is an important step \cite{OL31p1875,OE15p6677} 
toward determining the ultimate accuracy of femtosecond lasers as 
optical atomic clocks.

\ack

I thank Mark Ablowitz, Steven Cundiff, William Kath, 
Curtis Menyuk, Avner Peleg and Elaine Spiller for many 
interesting discussions.
This work was supported by NSF under grant number DMS-0506101.

\section*{Appendix:  Numerical methods for the DMGLE}
\setcounter{section}1
\renewcommand\theequation{\Alph{section}.\arabic{equation}}

Two relevant issues are:
(i)~efficient methods for numerical integration of the time-dependent
DMGLE~\eref{e:DMGLE},
and (ii)~numerical methods to find dispersion-managed soliton solutions, 
i.e., solutions of the nonlinear integral equation~\eref{e:DMSinteqn}.
The first issue is rather straightforward:
equation~\eref{e:DMGLE} can be integrated using essentially 
the same techniques as for the DMNLSE.
These techniques include a method for the fast evaluation of
the double integral.
Since this method was discussed in detail in the appendix
of Ref.~\cite{PRA75p53818}, we do not repeat
that discussion here, and we turn our attention instead
to the nonlinear integral equation~\eref{e:DMSinteqn}.

When $\=b=\=c=\=g=0$ (that is, for the DMNLSE),
\eref{e:DMSinteqn} can be efficiently
solved numerically using Petviashvili's method
(as discussed in the Appendix of Ref.~\cite{PRA75p53818}).
Thus, here we only need to address the case when $\=b$, $\=c$ or 
$\=g$ are nonzero.
For simplicity, we first present the method for the CGLE 
(that is, when $s=0$),
even though in that case one can find stationary solutions analytically.
This will allow us to explain the relevant ideas ---
which are the same for both the CGLE and the DMGLE ---
without some of the notational complications resulting
from the presence of the kernel~$r(k'k'')$ in~\eref{e:DMSinteqn}.

As with the DMNLSE, we solve the nonlinear integral equation~\eref{e:DMSinteqn}
numerically by introducing an appropriate iteration scheme.
Compared to the DMNLSE, however,
two additional complications must be addressed:
\begin{enumerate}
\item
An appropriate correction factor is necessary for convergence, 
since a standard Neumann iteration diverges.
This issue is similar to that arising for the DMNLSE.
In that case, the problem can be solved using Petviashvili's method.
That method does not converge for the CGLE/DMGLE, however, 
due to the presence of both a real and an imaginary component in 
the solution.
\item
The propagation constant $\mu$ is unknown a priori. 
That is, unlike the NLSE/DMNLSE, soliton solutions of the 
CGLE and DMGLE only exist for certain discrete values of 
the propagation constant, cf.~\eref{e:CGLEsoliton} and~\eref{e:c}.
Since these values are not known in advance,
in addition to looking for the functions $F\_r$ and $F\_i$
the solution method must simultaneously look for the
values of $\mu$ for which~\eref{A:F} admits nontrivial solutions.
\end{enumerate}
We discuss both problems below.

We seek a complex-valued function $f(x)= f_\re+i f_\im$ 
(with $f_\re$ and $f_\im$ both even functions of~$x$) 
and a propagation constant $\mu=\lambda^2/2$ 
such that $F(k):= \F[f]$ 
satisfies~\eref{e:DMSinteqn}, 
which, when $s=0$ (that is, for the CGLE) reduces to
\[
\big(\mu+\half(\=d-i\=b)k^2\big)\,\^f+ i\=g \^f = (1+i\=c)\F[fff^*]\,.
\label{A:F}
\]
Decomposing~\eref{A:F} into its real and imaginary parts yields the 
following two-component real system of nonlinear integral equations:
\bea
(\~A+\mu\,\~I)\,\@F - \~C\,\@N(\@F) = 0\,,
\label{A:AF=CN}
\\
\noalign{\noindent where}
\@F= \left(\!\!\begin{array}{c}
  \^f_\re\\\^f_\im\end{array}\!\!\right)\,,\qquad
\@N(\@F)= \left(\!\!\begin{array}{c}
  \F[f_\re^3]+\F[f_\re f_\im^2]\\
  \F[f_\re^2f_\im]+\F[f_\im^3]
\end{array}\!\!\right)\,,
\label{A:NrNi}
\\
\noalign{\noindent $\~I$ is the $2\times2$ identity matrix
and}
\nonumber
\\[-2ex]
\~A(k) = \left(\!\!\begin{array}{cc}\half\=dk^2 &\half\=bk^2-\=g\\
  \=g-\half\=bk^2 &\half\=dk^2\end{array}\!\!\right)\,,
\qquad
\~C = \left(\!\!\begin{array}{cc}1&-\!\=c\\\=c&1\end{array}\!\!\right)\,.
\eea
(Of course $\^f_\re$ and $\^f_\im$ are also both even functions of~$k$.)

Let us first address the issue of the unknown propagation constant.
Suppose that $\mu_o$ is the exact propagation constant,
and $\mu= \mu_o + \Delta\mu$.
Equation~\eref{A:AF=CN} then yields
\[
\Delta\mu\,\@F = \~C\,\@N(\@F)- (\~A+\mu_o\,\~I)\,\@F\,.
\label{e:muexcess}
\]
We can therefore can obtain $\Delta\mu$ as
\bea
\Delta\mu = \frac{ \<\@F,\~C\,\@N(\@F)\> - \<\@F,\~A\,\@F\>}
  {\<\@F,\@F\>}-\mu_o\,,
\label{A:deltamu}
\eea
Here the inner product of two real vector functions is
\bea
\<\@F,\@G\>= \int \@F(k)\cdot\@G(k)\,\d k\,.
\eea
Of course~\eref{A:deltamu} contains the solution $\@F$, 
which is not known exactly during the iteration.
Nonetheless, \eref{A:deltamu} provides a way to update our
estimate of the correct eigenvalue.

We now turn to the issue of convergence factors.
This problem can be dealt with using the 
spectral renormalization method~\cite{OL30p2140}.
To implement this method, we start by noting that 
$\mu$ is not sign-definite, and even in those cases where $\mu$
is positive, some iterations might accidentally yield
a negative estimate.
In order to avoid problems with zero denominators, it is 
therefore convenient to add and subtract the term $\~I\,\@F$ 
from equation~\eref{A:AF=CN}, rewriting it as:
\[
(\~A+\~I)\,\@F= (1-\mu)\,\@F + \~C\,\@N(\@F)\,,
\label{A:F=MN}
\]
\renewcommand\theequation{\Alph{section}.\arabic{equation}}%
We then introduce the real convergence factor $\gamma$
and rescale the solution as $\@F= \gamma\,\@V\,$.
Noting that $\@N(\@F)= \gamma^3 \@N(\@V)$, for the new field 
we obtain the system
\[
\@V= (\~A+\~I)^{-1}\big[\,(1-\mu)\,\@V + \gamma^2\~C\,\@N(\@V)\,\big]\,,
\label{A:V=MN}
\]
which provides the basis for the iteration.
An equations for the convergence factor is also needed, of course.
This is obtained by using~\eref{A:F=MN} with $\@F=\gamma\@V$
and taking its inner product with $\@V$, 
which yields
\bea
\gamma^2= 
  \<\@V,(\~A + \mu\,\~I)\@V\>\big/\<\@V,\~C\,\@N(\@V)\>\,.
\label{A:conv}
\eea

Combining the two results, 
we can then define an iteration scheme as follows.
At $n=0$, choose an initial guess for $f_\re\O0+if_\im\O0$ and $\mu\O0$,
and set $\gamma\O0=1$ and $\@V\O0= \@F\O0$.
Then, at the $(n+1)$-st step of the iteration:
\begin{enumerate}
\def\labelenumi{\arabic{enumi}.}
\item
Update the convergence factor $\gamma\O{n}$ 
and the current estimate of the eigenvalue $\mu\O{n}$
using~\eref{A:conv} 
[with $\@V= \@V\O{n}$ and $\mu=\mu\O{n}$]
and~\eref{A:deltamu}
[with $\@F= \gamma\O{n}\@V\O{n}$ 
and $\mu=\mu\O{n}$].
\item
Update the current estimate of the solution using~\eref{A:V=MN}
with the newly updated values for $\gamma\O{n}$ and $\mu\O{n}$.
\end{enumerate}
At each iteration, the nonlinear terms can be computed 
using the fast Fourier transform (FFT), which reduces the
computational cost of each step from from $O(N^3)$ to $O(N^2\log N)$,
where $N$ is the number of grid points or Fourier modes.

The method to find soliton solutions of the DMGLE works exactly in 
the same way as that for the CGLE above.
The only difference is that the term $\F[fff^*]$ in the RHS of~\eref{A:F} 
is replaced by the double convolution integral $P[\^f,\^f,\^f^*,r]$, where
\bea
P[\^f_1\^f_2,\^f_3^*,r]=
  \iint \^f_1{(k+k')}\^f_2{(k+k'')}\^f_3^*{(k+k'+k'')}r(k'k'')\d k'\d k''\,.
\nonumber
\eea
The nonlinear term $\@N(\@F)$ is modified accordingly.
Apart from these changes, the method is exactly the same as for the CGLE.
Moreover, as in the DMNLSE,
the resulting double convolution integrals can still be computed efficiently
using FFTs. 
As this issue was explained in detail in the appendix of Ref.~\cite{PRA75p53818},
we dot duplicate that discussion here.

\makeatletter
\def\note#1{~\textit{\small[#1]}}
\def\@biblabel#1{#1.}
\def\title#1{``#1''}
\def\booktitle#1{\textsl{#1}}
\def\journal#1&#2,#3(#4){\begingroup{\sl #1\unskip}\ {\bf\ignorespaces #2}\rm, #3 (#4)\endgroup}

\end{document}